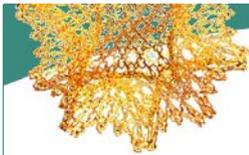
Mathematica Italia
User Group Meeting

# Making simple proofs simpler

Pietro Codara, Ottavio D'Antona, Francesco Marigo, Corrado Monti

Dipartimento di Informatica, Università degli Studi di Milano

## Abstract

An open partition $\pi$ [Cod09a, Cod09b] of a tree $T$ is a partition of the vertices of $T$ with the property that, for each block $B$ of $\pi$, the upset of $B$ is a union of blocks of $\pi$. This paper deals with the number, NP($n$), of open partitions of the tree, $V_n$, made of two chains with $n$ points each, that share the root. We prove that the number of such partitions is given by the formula

$$\mathrm{NP}(n) = \sum_{k=1}^{n}\sum_{j=1}^{n}\binom{n-1}{k-1}\binom{n-1}{j-1}\mathrm{Min}[k,j]$$

Below are a few values of NP($n$):

$$1, 5, 26, 130, 628, 2954, 13612, 61716, \ldots$$

On the other hand, from the On-Line Encyclopedia of Integer Sequences [OEIS] we learn that such a numerical sequence is generated by the formula

$$(n+1)\,2^{2n-3} - \frac{n-1}{2}\binom{2n-2}{n-1}$$

In this note we exploit Mathematica to explain the not so simple derivation of the following novel identity, that will prove the equivalence between the above formulas:

$$\mathrm{NP}(n) = \sum_{k=1}^{n}\sum_{j=1}^{n}\binom{n-1}{k-1}\binom{n-1}{j-1}\mathrm{Min}[k,j] = (n+1)\,2^{2n-3} - \sum_{k=1}^{n}\sum_{h=1}^{k-1}(k-h)\binom{n-1}{h-1}\binom{n-1}{k-1}$$

In so doing we show that visual enumeration can make simple proofs simpler!

# 1. Introduction

In our research we often deal with combinatorial problems, sometimes quite complex, on ordered structures. Recall that a partially ordered set (poset, for short) is a set together with a reflexive, transitive, and antisymmetric binary relation (usually denoted by $\leqslant$) on the elements of the set.

Posets are usually represented as a Hasse diagram, as in [Cod10b]; that is, the bigger elements (according to $\leqslant$) are drawn on the top, and line starting in the element x connects it to all its direct successors (any z for which $z \leqslant x$ but there are not y such that $z \leqslant y \leqslant x$). If $z \leqslant y$ and $y \leqslant x$, nodes are drawn in this order, from bottom to top, in the Hasse diagram. Uncomparable nodes (x,y such that neither $x \leqslant y$ nor $y \leqslant x$) are not connected by a bottom-to-top path, neither with nor without nodes in the middle.

Our attention is mainly addressed to a certain category of posets: those composed by two chains that share the same root. The Hasse diagram for this kind of structures is this:

In[21]:=
```
n =3;
catL = Join[{1},Range[2, n*2-1,2]];
catR = Range[1, n*2-1,2];
nodes =Union[catL,catR];
poset= MakeGraph[nodes,IsEdge];
PosetPlot[poset]
```

Out[26]=

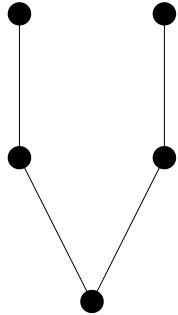

This will allow us to call a structure of this kind a ***V-Poset***.

Now, we are interested in partitions of a poset: namely, a **partition** of a poset is a division of its nodes into non-overlapping and non-empty **blocks** whose union is the set of nodes. There are many ways to define classes of possible partitions on posets that are useful to conserve some property about the relationship that defines the poset. In particular, we will use **open partitions** [Cod09a, Cod09b]: an open partition of a Poset is, informally speaking, a partition where, for any block X, every node that is under a node included in X is part of a block where all the nodes are under an element of X.

We can formally describe this property using the notion of filter. A filter of an element x of a Poset $(P,\leqslant)$ is a set, denoted as $\uparrow x$, of all the elements $y \in P$ such that $y \leqslant x$. We can now associate to a partition $\pi$ another relationship $\preccurlyeq$ such that $B \preccurlyeq C$ for $B,C \in \pi$ if $\exists\, b \in B$ and $\exists\, c \in C$ for which $b \leqslant c$. Now, if $\pi$ is an open partition, for each of its blocks B exists a collection $\kappa \subseteq \pi$ of blocks of $\pi$ such that

$$\uparrow B = \bigcup_{C \in \kappa} C$$

This can be easily decribed in this way: the filter of a block does not "break" any other block.

Partitions will be drawn in this work automatically by a simple Mathematica package we developed, by means of colors: elements in the same block will have same color; those in different blocks will have different colors. We remind the reader that, since a block of a partition is defined only by its member, the color is not important: given a partition, and swapping all the colors of its blocks, we will have exactly the same partition; no matter what the color of its blocks are, that will not be a different partition.

Now let us see how an open partition of a V-poset look like:

```
aPartition={{1,2},{3}, {4}, {5}};
PosetPlot[poset, aPartition]
```

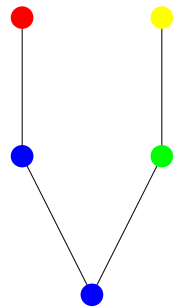

We can see how under the red block we have the blue one, but no node of the other blocks; in the same way, under the yellow block we have the green one and the blue one, but no other elements for the other blocks.

# 3. How many open partitions could we have?

We are interested in the following combinatorial problem: given a V-poset composed by two chains of *n* elements each, how many open partitions could we have? Is there a closed form? Can this be expressed without explicit summations?

To count these kind of partitions, it is convenient to define them in a simpler way. Indeed, we can note that they are completely defined by:

- a partition on the left chain
- a partition on the right chain
- a number of blocks to join, at the bottom, between the two partitions (we necessarily have to join the bottom block of both, because they share the root node)

Let us see this with *Mathematica*:

> ***Note:*** to see the Manipulating boxes, **please execute all Initialization Cells**:
>
> Menu→Evalutation→Evaluate Initialization Cells

Out[30]//TableForm=

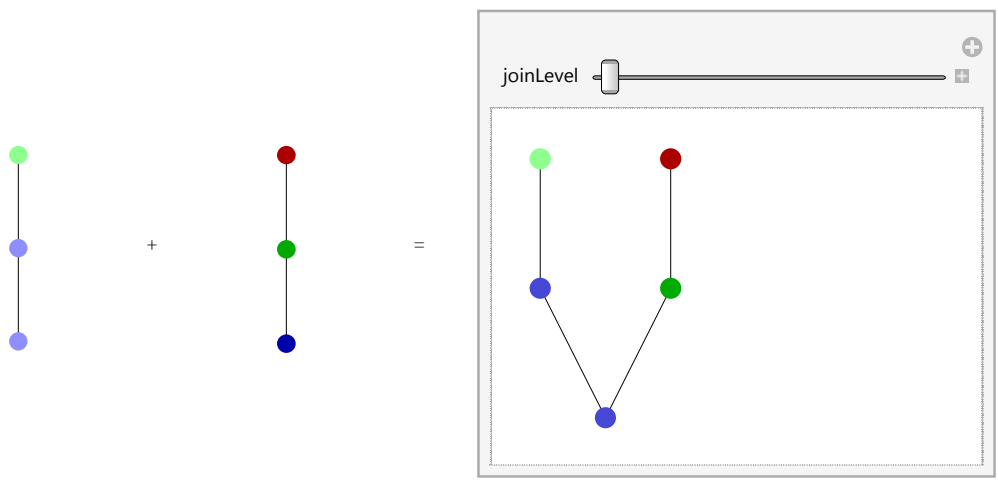

This provides naturally a counting of the partitions on the V-poset. All we need to know is that the number of open partitions on a chain of *n* elements in *k* block is $\binom{n-1}{k-1}$ (to briefly prove this fact, just think as putting k-1 "separators" on the edges of the chain). Therefore, we can count all the possible partitions on each chain with *k* and *j* blocks respectively, and we know that they can be joined in *Min{ k , j }* different ways. Calling *n* the number of nodes in each chain, this gives us the following formula:

$$\mathbf{NP}(n) = \sum_{k=1}^{n}\sum_{j=1}^{n}\binom{n-1}{k-1}\binom{n-1}{j-1}\mathbf{Min}[k, j]$$

With *Mathematica*, we can illustrate the meaning of this:

Out[31]=
$$\sum_{k=1}^{n}\sum_{j=1}^{n}\binom{n-1}{k-1}\binom{n-1}{j-1}\text{Min}[k, j]$$

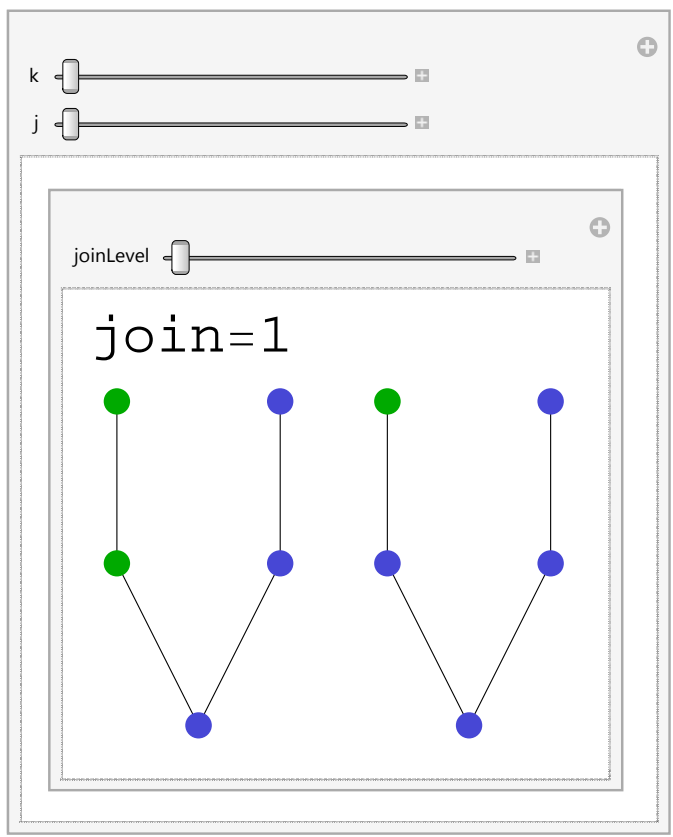

Out[32]=

# 4. Our first statement

$$\text{NP}(n) = \sum_{k=1}^{n}\sum_{j=1}^{n}\binom{n-1}{k-1}\binom{n-1}{j-1}\text{Min}[k,j] =$$

$$= \sum_{j=1}^{n}\left(\sum_{k=j}^{n}\binom{n-1}{k-1}\right)^2$$

### 4.1 Algebraic proof

This first statement is simple but not exactly trivial. It can be proved with algebra; we also tried to do it symbolically with *Mathematica* but effort was of little avail. First,

$$\text{NP(n)} = \sum_{k=1}^{n}\sum_{h=1}^{n}\text{Min}[k,h]\cdot\binom{n-1}{k-1}\binom{n-1}{h-1} =$$
$$= \sum_{j=1}^{n}\sum_{k=j}^{n}\sum_{h=j}^{n}\binom{n-1}{k-1}\binom{n-1}{h-1}$$

since every couple *k,h* is counted exactly Min[*k,h*] times if the index of the inner summation starts with *j*. Now, let us move the first binomial coefficient out:

$$= \sum_{j=1}^{n}\sum_{k=j}^{n}\left[\binom{n-1}{k-1}\sum_{h=j}^{n}\binom{n-1}{h-1}\right] =$$

and then, also the inner sum:

$$= \sum_{j=1}^{n} \left( \sum_{h=j}^{n} \binom{n-1}{h-1} \right) \left( \sum_{k=j}^{n} \binom{n-1}{k-1} \right) =$$

In the end, we have:

$$= \sum_{i=1}^{n} \left[ \sum_{j=i}^{n} \binom{n-1}{k-1} \right]^2$$

Which was what we wanted.

### 4.2 A visual proof

Nevertheless, we know the meaning of the first formula, and we have visualized it. We could prove the equivalence in a combinatorial way, just by understanding why the second formula counts exactly the same objects counted by the first one. Proving the fact that the two expressions count the same objects is the same as proving that they are equal. With *Mathematica* we can visualize these summatories and show explicitly how the objects that are counted are, again, all the open partitions of a V-poset:

Out[33]=
$$\sum_{j=1}^{n} \left( \sum_{k=j}^{n} \binom{n-1}{k-1} \right)^2$$

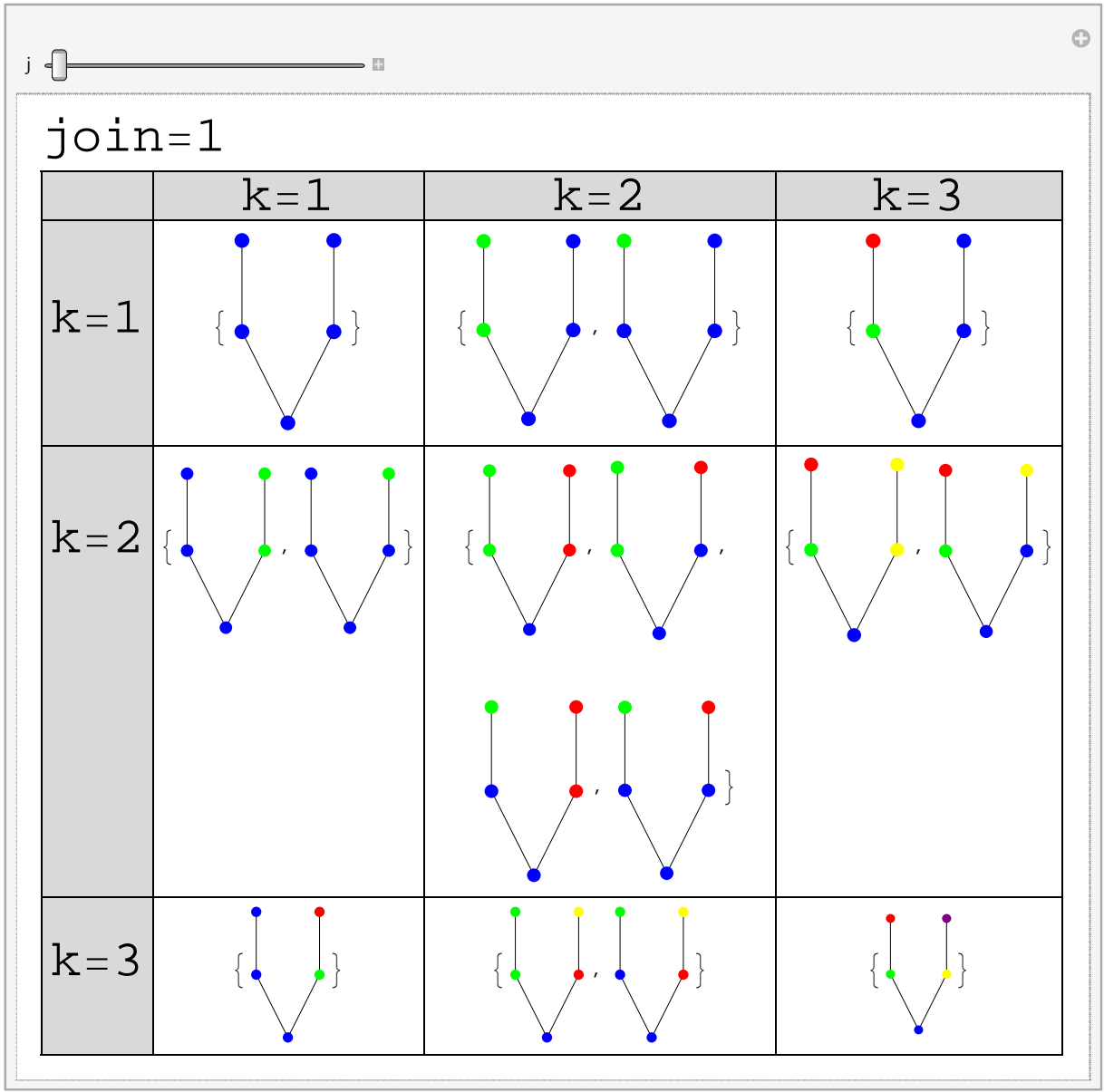

# 5. A second statement

$$NP(n) = 2^{n-1} * (n+1) \, 2^{n-2} - \sum_{k=1}^{n} \sum_{h=1}^{k-1} (k-h) \binom{n-1}{h-1} \binom{n-1}{k-1}$$

### 5.1 Algebric proof

This proof is not simple in algebraic ways.
We start from the equation

$$NP(n) = \sum_{h,k=1}^{n} min(h,k) \binom{n-1}{h-1} \binom{n-1}{k-1}$$

In order to give another expression to *NP(n)*, we use the following lemma:

**Lemma 1.**
Let $a_i$, $b_i$ with *i=1,...,n* be two symmetric sequences of positive real numbers, that is $a_i = a_{n+1-i}$ and  $b_i = b_{n+1-i}$.

Let $A = \sum_{i=1}^{n} a_i$ and $B = \sum_{i=1}^{n} b_i$ be the sums of the two sequences. Then

$$\sum_{h,k=1}^{n} min(h, k)\, a_h\, b_k = \frac{n+1}{2} A\, B - \sum_{h \leq k=1}^{n} (k - h)\, a_h\, b_k$$

*Proof.*
Let us write

$$X = \sum_{h,k=1}^{n} min(h, k)\, a_h\, b_k$$

We can change the indexes into $h \to n + 1 - h$ and $k \to n + 1 - k$, and, because of the simmetry of the two sequences, we have

$$X = \sum_{h,k=1}^{n} (n + 1 - max(h, k))\, a_h\, b_k$$

Summing the two expressions for *X*, we have

$$2X = \sum_{h,k=1}^{n} (n + 1 + min(h, k) - max(h, k))\, a_h\, b_k =$$
$$= \sum_{h,k=1}^{n} (n + 1 - |k - h|)\, a_h\, b_k =$$
$$= (n + 1) \sum_{h,k=1}^{n} a_h\, b_k - \sum_{h,k=1}^{n} |k - h|\, a_h\, b_k =$$
$$= (n + 1) A\, B - 2 \sum_{h \leq k=1}^{n} (k - h)\, a_h\, b_k$$

And dividing by two we have

$$X = \frac{n+1}{2} A\, B - \sum_{h \leq k=1}^{n} (k - h)\, a_h\, b_k,$$

This proves the lemma.
□

This result is also very similar to the one proved by Codara in [Cod10a]. Now we apply this lemma to our equation, with the sequences

$$a_i = b_i = \binom{n-1}{i-1}$$

We have
$$NP(n) = \sum_{h,k=1}^{n} min(h, k) \binom{n-1}{h-1}\binom{n-1}{k-1} =$$
$$= \frac{n+1}{2} \sum_{h=1}^{n} \binom{n-1}{h-1} \sum_{k=1}^{n} \binom{n-1}{k-1} - \sum_{h \leq k=1}^{n} (k - h) \binom{n-1}{h-1}\binom{n-1}{k-1} =$$
$$= \frac{n+1}{2} 2^{n-1}\, 2^{n-1} - \sum_{h \leq k=1}^{n} (k - h) \binom{n-1}{h-1}\binom{n-1}{k-1}$$
and finally
$$NP(n) = (n + 1)\, 2^{2n-3} - \sum_{h \leq k=1}^{n} (k - h) \binom{n-1}{h-1}\binom{n-1}{k-1}$$
which was what we wanted.

## 5.2 Visual proof

Again, to prove the statement in a combinatorial way, we just have to show that it counts all the open partitions of a V-poset. This time, though, we do not have directly a summation, but the formula is composed by a difference instead. So, we will recognize what every part of this difference means for our partitions.

The first part — the minuend — is a product. A product, in combinatorics, very often means a cartesian product between two sets: this is the case. In fact, this product can be seen as the product between these two sets. The following, of $2^{n-1}$ elements, is the set of all the open partitions of the left chain of the V-poset:

Out[35]=
$$2^{n-1}$$

Out[39]=
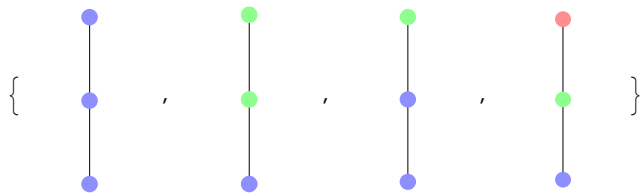

The equivalence between the sommation of binomials with the same top number and the corrispondent power of 2 is indeed well known.

The second set of the cartesian product is made by all the pairings of an open partition on the right chain, and a number that can be seen as a join level. This number will range from 1 to the number of blocks defined by the partition on this chain. So we have these objects:

Out[40]=
$$(n+1)\, 2^{n-2} \;=\; 2^{n-1} \cdot \frac{n+1}{2}$$

Out[42]=
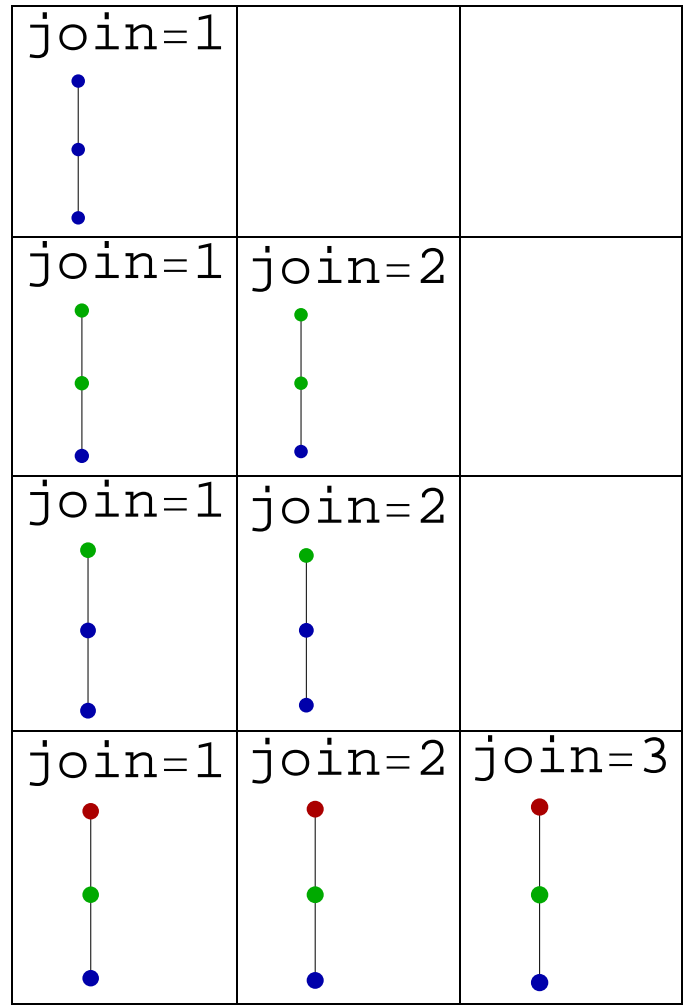

The trick to understand the subtraction of our formula is that not all the pairings between these two sets are legal. What is counted by the subtrahend of the formula are extacly these uncorrect pairings. Let us see this directly:

$$\sum_{k=1}^{n}\sum_{h=1}^{k-1} (k-h) \binom{n-1}{h-1} \binom{n-1}{k-1}$$

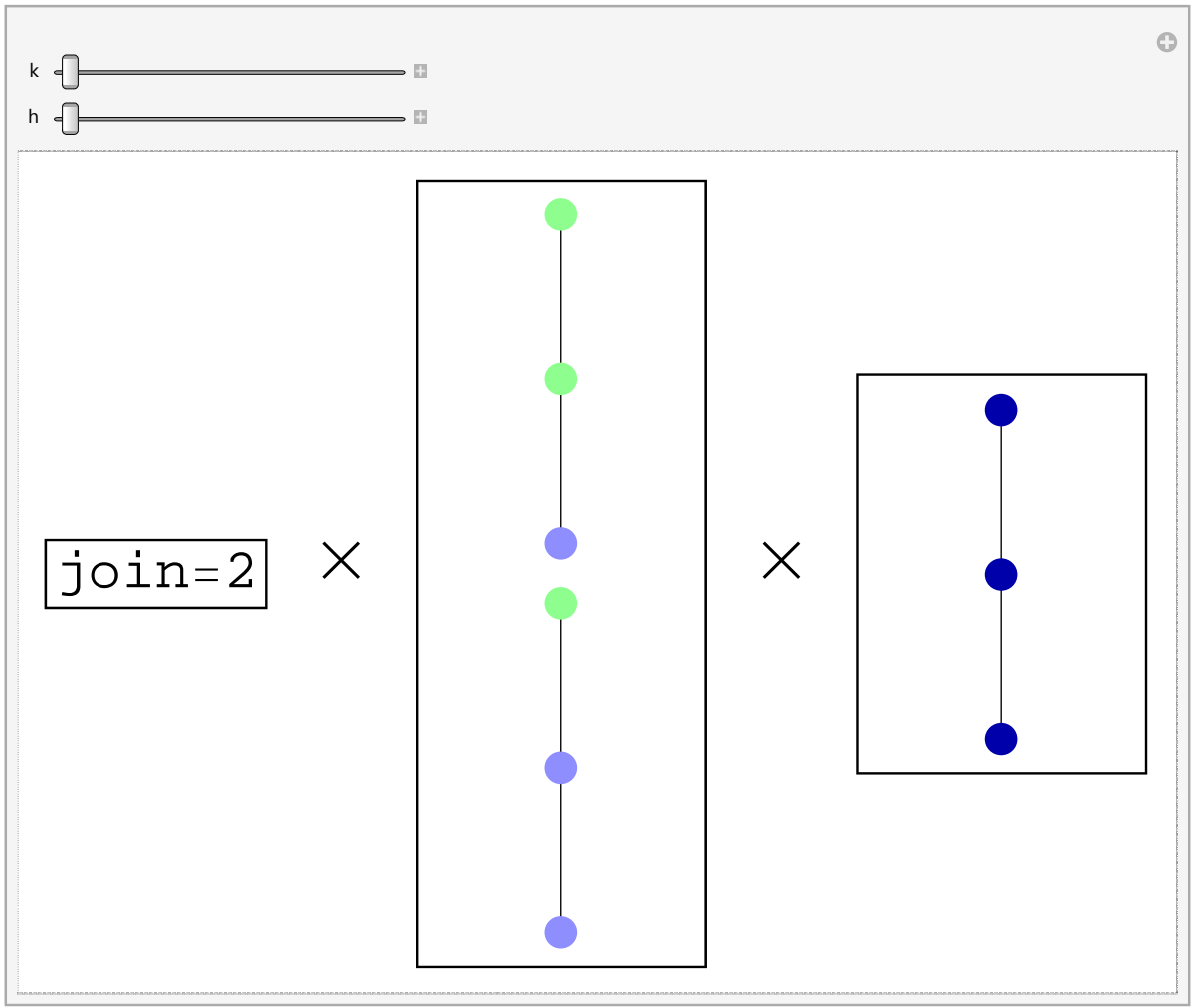

So our formula can be seen as a subtraction of elements from a product, and can be represented in this way:

$$2^{n-1} * (n+1)\, 2^{n-2} - \sum_{k=1}^{n} \sum_{h=1}^{k-1} (k-h) \binom{n-1}{h-1}\binom{n-1}{k-1}$$

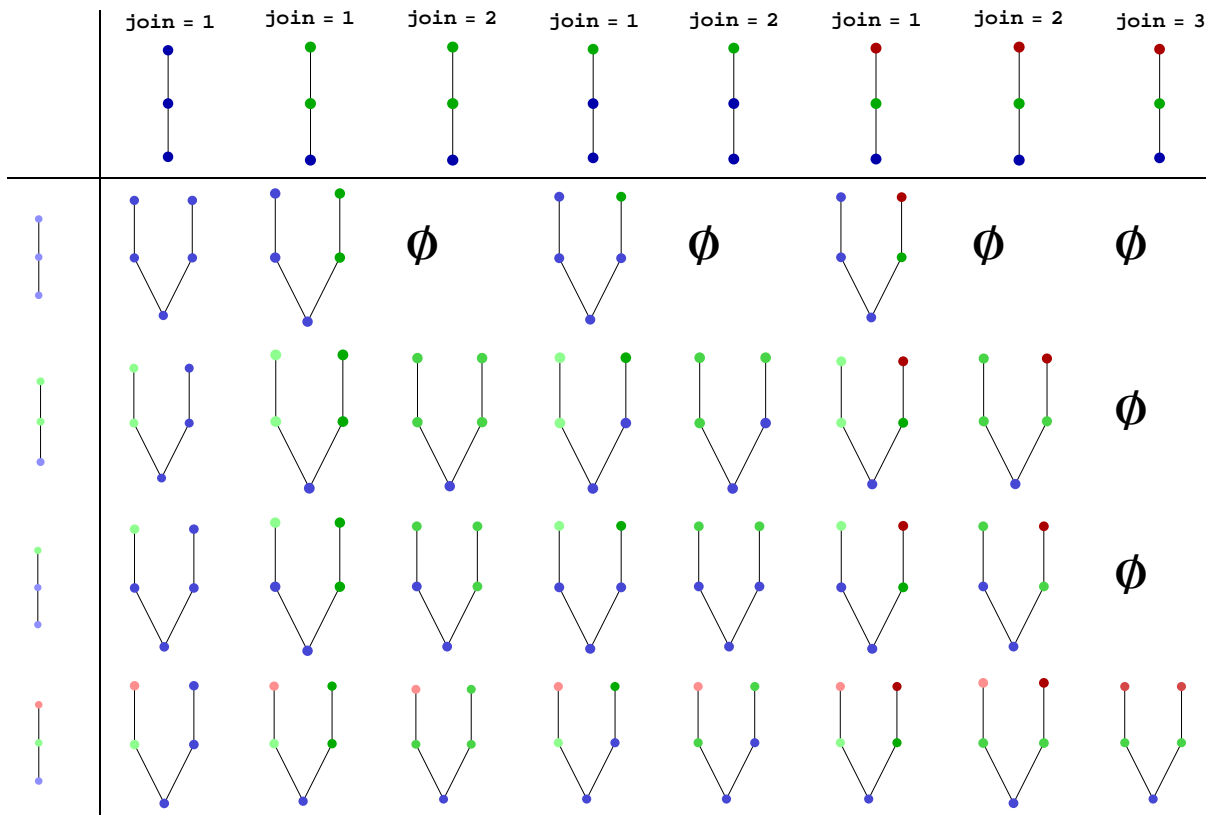

Proving that the above formula counts all the elements of the first one we have indirectly proved the equation itself, without all the algebraic machinery otherwise involved.

Moreover, thanks to a result by Hirchhorn [Hirsch96] that says that

$$\sum_{k=1}^{n} \sum_{h=1}^{k-1} (k-h) \binom{n-1}{h-1} \binom{n-1}{k-1} = \frac{n-1}{2} \binom{2n-2}{n-1}$$

we can build — on the top of our result — this derivative form:

$$\text{NP}(n) = (n+1) 2^{2n-3} - \frac{n-1}{2} \binom{2n-2}{n-1}$$

That is a formula for P that avoid explicits summation (and, therefore, is computationally cheaper). Additional results — like the generalization of this new form to the case where the two chains differ in length — can be also proved without much effort, using the interpretation we have (literally!) seen.